\documentclass[preprint,12pt,authoryear]{elsarticle}

\usepackage{amssymb}
\usepackage{amsmath}
\usepackage[figuresright]{rotating}
\usepackage{algorithm}
\usepackage{algpseudocode}
\usepackage[export]{adjustbox}
\usepackage{setspace} 
\usepackage{mathrsfs}
\usepackage{multirow}
\usepackage{subfigure}
\usepackage{rotating}
\usepackage{booktabs}
\usepackage{subcaption}
\usepackage{graphicx}
\usepackage[table]{xcolor}
\usepackage{tabularx}
\usepackage{caption}
\usepackage{bm}
\usepackage{url}

\makeatletter
\def\ps@pprintTitle{%
  \let\@oddhead\@empty
  \let\@evenhead\@empty
  \let\@oddfoot\@empty
  \let\@evenfoot\@oddfoot
}
\makeatother

\begin{document}

\begin{frontmatter}



\title{Composite method for fast computation of individual level spatial epidemic models} 


\author[affa]{Yirao Zhang\corref{cor1}} 
\author[affa]{Rob Deardon\fnref{affb}} 
\author[affc]{Lorna Deeth} 
\cortext[cor1]{Corresponding author. Email address: yirao.zhang1@ucalgary.ca}
\affiliation[affa]{organization={Department of Mathematics and Statistics, University of Calgary},
            addressline={2500 University Dr NW}, 
            city={Calgary},
            postcode={T2N 1N4}, 
            state={AB},
            country={Canada}}
\affiliation[affb]{organization={Faculty of Veterinary Medicine, University of Calgary},
            addressline={3280 Hospital Dr NW}, 
            city={Calgary},
            postcode={T2N 4Z6}, 
            state={AB},
            country={Canada}}
\affiliation[affc]{organization={Department of Mathematics and Statistics, University of Guelph},
            addressline={50 Stone Road East}, 
            city={Guelph},
            postcode={N1G 2W1}, 
            state={ON},
            country={Canada}}

\begin{abstract}
Individual-level models, also known as ILMs, are commonly used in epidemics modelling, as they can flexibly incorporate individual-level covariates that influence susceptibility and transmissibility upon infection. However, inference for ILMs is computationally intensive, especially as the total population size increases and additional covariates are incorporated. We propose a composite method, the composite ILM (C-ILM), that clusters the population into minimally-interfered subpopulations, with between-cluster infections enabled through a ``spark function.'' This approach allows for parallel computation of subsets before aggregation. Focusing on C-ILM, we consider four ``spark functions'', and introduce a Dirichlet process mixture modelling (DPMM) algorithm for clustering. 
Simulation results indicate that, in addition to faster computation, C-ILM performs well in parameter estimation and posterior predictions. Furthermore, within C-ILM framework, DPMM algorithm demonstrates superior performance compared to the conventional $K$-means algorithm. We apply the methods to data from the 2001 UK foot-and-mouth disease outbreak. The results provide evidence that C-ILM is not only computationally efficient but also achieves a better model fit compared to the basic spatial ILM.
\end{abstract}




\end{frontmatter}



\section{Introduction}
Infectious diseases are typically modelled using multi-state processes, such as the compartmental framework introduced by \citet{kermack1991contributions}. However, traditional compartmental models like the susceptible-infected-removed (SIR) model face limitations due to the assumption of population homogeneity, which often does not hold true. The individual-level model (ILM) framework proposed by \citet{deardon2010inference} addresses these issues within a compartmental framework. The ILM framework enables the incorporation of individual-level covariate information that may affect susceptibility to a disease. It also relaxes the homogeneous mixing assumption often made in population-level models by accounting for spatial distance between individuals or connections via a contact network. When fitted in a Bayesian context, analyses with ILMs can also accommodate missing data and incorporate prior information. Fitted ILMs facilitate the identification of potential risk factors and enhance understanding of infectious disease transmission processes. Moreover, they offer the potential to forecast the trajectory of ongoing outbreaks and assess intervention policies.\par
To capture even more intricate transmission dynamics, the ILM framework has been recently extended to accommodate spatial considerations at the regional level, via the geographically dependent ILM (GD-ILM) \citep{mahsin2022geographically}, and temporal variations in transmission dynamics attributable to behavioural change (BC) prompted by individual reaction and/or government-mandated protective measures (BC-ILM) \citep{ward2025framework}.  While extensions of the ILM framework improve its flexibility, they can also increase computational demands by introducing additional variables and/or structure. For example, GD-ILMs incorporate regional level random effects, and BC-ILMs introduce new variables to account for time-varying behavioural effects. In any case, under both default and enhanced spatial ILM frameworks, the computational load remains significant for large populations, primarily due to the necessity of considering each individual's non-linear and time-varying infection rate in the likelihood. To mitigate the computational intensity associated with likelihood calculations, various approaches have been introduced: \citet{pokharel2016gaussian} introduced an emulation-based inference approach, and later incorportated event time uncertainty into the framework \citep{pokharel2022emulation}; \citet{malik2016parameterizing} proposed a sampling-based approach followed by various spatially-stratified schemes; \citet{almutiry2020incorporating} proposed an approximate Bayesian computation (ABC) approach in the context of uncertain contact network data; and \citet{ward2022computationally} introduced an ``aggregate-disaggregate'' method that clustered data into aggregate units before model fitting and compared it to ABC approaches. However, these methods are often difficult to implement, involve tuning parameters, and generally lead to approximations of model/likelihood that may be difficult to interpret. \par
As another possible remedy, ILMs can be fitted using so-called divide-and-conquer methods, wherein cluster-based data subsets are independently analyzed in parallel before being aggregated \citep{almutiry2018incorporating}. One notable such method is Consensus Monte Carlo \citep{scott2022bayes}, which facilitates distributed approximate Bayesian analyses by executing separate Monte Carlo algorithms on individual machines and subsequently averaging Monte Carlo samples across these machines. Another communication-free parallel technique is the Likelihood Inflation Sampling Algorithm (LISA) \citep{entezari2018likelihood}, which similarly reduces computational expenses by partitioning data into smaller subsets and conducting MCMC sampling from inflated sub-posterior distributions using different processors in parallel. However, in their basic form, these methods assume data sets can be analyzed independently. Thus, when applied in an epidemic scenario, all these approaches in their naïve state preclude between-subset transmission, potentially impeding the accuracy of estimative and predictive inference. \par
The central objective of this study is to propose a novel framework of ILMs that allows for more efficient rapid inference: the composite ILM (C-ILM). Motivated by divide-and-conquer approaches, this method involves two stages. First, a clustering method is used to divide the population into spatial subpopulations. The goal here is to define subpopulations in which the within subpopulation infections occurs at a much higher rate than between subpopulation infections. The second stage consists of using the C-ILM to analyze the epidemic data conditional on these subpopulations. Between-cluster infections are either disallowed under the C-ILM, or facilitated via a mechanism involving a much lower computational burden than under the original ILM. This is done via the so-called ``spark function." This framework also potentially allows the likelihood to be calculated in parallel, significantly further reducing computation time.\par
In this study, we devise a method based upon Dirichlet process mixture modelling (DPMM) with a stick-breaking representation for clustering the population prior to fitting the C-ILM. This method utilizes information on individual spatial locations and infection times. We also compare this method to a simpler $K$-means clustering algorithm that utilizes only spatial locations to cluster the population. We explore various spark functions and the performance of the DPMM clustering algorithm for C-ILMs through a simulation study, and then demonstrate our methods using real-world data from the 2001 UK foot-and-mouth disease (FMD) epidemic in livestock. 
\par
In the following section, we introduce the general framework of ILMs and the composite methods for the spatial ILM, followed by the DPMM clustering algorithm in Section~\ref{s:dpmm}. The simulation study and its results are presented in Section~\ref{s:simulation}, and the application to the 2001 UK FMD data is discussed in Section~\ref{s:application}. Finally, in Section~\ref{s:discuss}, we discuss implications and future work.

\section{Composite methods for spatial individual level models (ILMs)}
\label{s:model}
\subsection{General Framework of ILMs}
\citet{deardon2010inference} introduced a framework for ILMs, which model the probability of infection over discrete time based on individual-level covariates and/or population mixing information. ILMs are placed within a compartmental (or multi-state) framework, such as the SIR framework we focus on here. In the SIR framework, at any given time point individuals in the population are categorized into one of three possible states: susceptible ($\mathcal{S}$), infectious ($\mathcal{I}$), or removed ($\mathcal{R}$). In the susceptible state, individuals are susceptible to the disease but have not been infected. In the infectious state, individuals are infected and capable of transmitting the disease to
others. In the removed state, individuals are assumed to have recovered from the disease with acquired immunity, been quarantined, or died, and are no longer able to contract or transmit the disease. State transitions occur in the direction: $\mathcal{S}\rightarrow\mathcal{I}\rightarrow\mathcal{R}$. The transition from susceptible to infectious is determined by the rate of infection, which in turn defines the infection probability. The transition from infectious to removed is determined
by the infectious period of the disease, which may be considered constant or can be allowed to vary between individuals. The general formulation of ILMs, proposed by \citet{deardon2010inference}, defines the probability of infection for individual $i$ at time $t$ as:
\begin{equation*}\label{general_form}
    P(i,t) = 1- \exp\left[\left\{-\Omega_{S}(i)\sum_{j\in I(t)}\Omega_{T}(j)\mathcal{K}(i,j)\right\}-\epsilon(i,t)\right],
\end{equation*}
where: $P(i,t)$ is the probability of a susceptible individual $i$ being infected at time $t$ (and becoming infectious at time $t+1$); $\Omega_S(i)$ is a susceptibility function of factors associated with the risk of  individual $i$ contracting the disease; $\Omega_T(j)$ is a transmissibility function of factors associated with the risk of individual $j$ passing on the disease; $I(t)$ denotes the set of infectious individuals at time $t$; $\mathcal{K}(i,j)$ is an infection kernel that accounts for risk factors involving both infectious and susceptible individuals, such as spatial separation (e.g., Euclidean or road distance) or edge weight in a contact network; and $\epsilon(i,t)$ is a spark term that captures infections not explained by $\Omega_S(i)$, $\Omega_T(j)$, and $\mathcal{K}(i,j)$, including infections originating from outside the population being observed.\par
Inference for ILMs is often conducted within a Bayesian Markov Chain Monte Carlo (MCMC) framework \citep{gibson1997markov}, which affords a powerful tool for carrying out inference on highly complex models and allows parameter uncertainty to be incorporated into forecasting and the testing of control strategies. The available epidemic data $\mathbf{D}$ typically consists of the infection times, removal times, spatial location, and any susceptibility and or transmissibility covariates, with associated parameters to be inferred ($\bm{\theta}$) depending on the scenario. Under an SIR-ILM framework, the likelihood for the epidemic data $\mathbf{D}$ across all time points in the study period is:
\begin{equation*}\label{full_lkhd}
    f(\mathbf{D}|\bm{\theta}) = \prod_{t=1}^{t_{\text{max}}-1}f_t(S(t), I(t), R(t)|\bm{\theta}),
\end{equation*}
where the likelihood at time $t$ can be written as: 
\begin{equation*}\label{t_lkhd}
    f_t(S(t), I(t), R(t)|\bm{\theta})=\left\{\prod_{i\in I(t+1)\backslash I(t)}P(i,t)\right\}\left\{\prod_{i\in S(t+1)}(1-P(i,t))\right\},
\end{equation*}
and where $I(t+1)\backslash I(t)$ is the set of newly infectious individuals at time $t+1$, and $S(t+1)$ is the set of the susceptible individuals at time $t+1$. $t_{\text{max}}$ is the last observed time point.\par
We initially focus on a simple spatial ILM (SILM) with no susceptibility and transmissbility covariates, in which the infection kernel is a power-law function of the Euclidean distance between susceptible and infectious individuals. Specifically, the susceptibility function is $\Omega_S(i) = \alpha$, the transmissibility function is $\Omega_T(j) = 1$, and the spatial infection kernel is $\mathcal{K}(i,j) = d_{ij}^{-\beta}$. The spatial parameter $\beta$ describes the decay of infection risk over increasing distance. The basic SILM is therefore given by
\begin{equation}\label{eq:basicSILM}
    P(i,t) = 1-\exp\left[\left\{ -\alpha \sum_{j\in I(t)} d_{ij}^{-\beta}\right\}-\epsilon(i,t)\right],\ \alpha,\beta>0.
\end{equation}
\subsection{Composite ILMs}
Here, we propose a novel composite modelling framework to accelerate inference for ILMs. Under the composite method, the population is partitioned into $K$ minimally interfering subsets. This can be done using simple methods such as $K$-means clustering of the spatial locations, or as we consider here, spatio-temporally defined using the DPMM clustering method introduced in Section \ref{s:dpmm}. 
A composite ILM (C-ILM) framework is then used in which within-cluster infections occur through the spatial infection kernel, while between-cluster infections are facilitated by the spark function (or indeed disallowed completely) in a computationally low-cost manner. The infection probability of individual $i$ in cluster $k$ under the general C-ILM is given by 
\begin{equation*}
P(i,t) = 1-\exp\left[\left\{-\Omega_{S}(i)\sum_{j\in I_{tk}}\Omega_{T}(j)\mathcal{K}(i,j)\right\}-\epsilon(i,t)\right], 
\end{equation*}
where $I_{tk}$ is the set of infectious individuals within cluster $k$ at time $t$. The simple spatial C-ILM is therefore given by 
\begin{equation}\label{simplespatial}
    P(i,t)=1-\exp\left[\left\{-\alpha\sum_{j\in I_{tk}}d_{ij}^{-\beta}\right\} - \epsilon(i, t)\right], \alpha,\beta>0.
\end{equation}
In the above equation~\ref{simplespatial}, the within-cluster infectivity rate is given by $\alpha\sum_{j\in I_{tk}}d_{ij}^{-\beta}$. The spark function $\epsilon(i, t)$ can now be used to account for between-cluster interference mechanisms. In Section~\ref{ss:spark}, various forms of $\epsilon(i, t)$ are considered. \par
We can broadly see that C-ILM can reduce time taken to compute the likelihood function. For example, assume the spark function computation time can be ignored as it is negligible compared to the computation time of the kernel. In the basic SILM, the total computation time is proportional to $\sum_{t=1}^{t_{\max}} N_{S_t} N_{I_t}$, where $N_{S_t}$ and $N_{I_t}$ are the number of susceptible and infectious individuals at time $t$, respectively. Using the composite method, the computation time for each cluster $k$ is proportional to $\sum_{t=1}^{t_{\max}} N_{S_{t,k}}N_{I_{t,k}}$, where $N_{S_{t,k}}$ and $N_{I_{t,k}}$ are the number of susceptible and infectious individuals at time $t$ in cluster $k$. Across all clusters, the total computation time is proportional to $\sum_{t=1}^{t_{\max}}\sum_{k=1}^{K} N_{S_{t,k}} N_{I_{t,k}}$. When the susceptible and infectious individuals are equally distributed over the clusters, the computation time can be reduced to $\sum_{t=1}^{t_{\max}}\frac{N_{S_t}N_{I_t}}{K}$. When $K$ is appropriately large, the computation time can be significantly reduced, although if $K$ is too large, or clusters are poorly chosen, there will likely be a loss of inferential performance.

\subsection{Spark functions}
\label{ss:spark}
There are many forms of the spark function that can be considered to allow for the between-cluster infections. Here, we consider five possibilities. To start, we consider the extreme case where $\epsilon(i, t) = 0$, which indicates between-cluster infections are not possible. To fit the model to data, at least one initial infection per cluster must be identified on which each cluster's component sub-likelihood is conditioned. Alternatively, we could employ a constant spark function term, $\epsilon(i, t) = \epsilon > 0$ and then conditioning on an infection per cluster is no longer necessary, as the spark function ensures a non-zero probability of ``spontaneous infection", facilitating disease spread. \par
A more complex scenario is to consider the number of infectious individuals in each cluster, along with the between-cluster distance. The spark term can be formulated as: $$\epsilon(i, t) = \epsilon \sum_{k^{'}\in K\backslash k} |I_{tk^{'}}|d_{kk^{'}}^{-\tilde{\beta}},\ \epsilon,\tilde{\beta}>0,$$
where: $K\backslash k$ denotes the set of all clusters except the one ($k$) containing individual $i$; $|I_{tk^{'}}|$ represents the number of infectious individuals in cluster $k^{'}$ at time $t$; and $d_{kk^{'}}$ is the distance between the centroids of clusters $k$ and $k^{'}$. The addition of a new parameter $\tilde{\beta}$ allows the between-cluster spatial effect in the spark function to differ from the within-cluster spatial effect in the kernel function. The C-ILM including this spark function is denoted as C-ILM M1 in Web Table~1.\par
The between-cluster infections can also depend on the susceptibility level of individual $i$, which in the basic SILM case is $\Omega_{S}(i) = \alpha$, yielding a spark function of: $$\epsilon(i, t) = \alpha\sum_{k^{'}\in K\backslash k} |I_{tk^{'}}|d_{kk^{'}}^{-\tilde{\beta}},\ \alpha,\tilde{\beta}>0,$$
which is C-ILM M2 in Web Table~1. By replacing $\epsilon$ with the susceptibility level $\alpha$, the number of parameters to be estimated is reduced.\par
Additionally, the distance term can be refined to consider the distance between the centroids of infectious individuals in cluster $k$ and $k^{'}$, leading a spark function of: $$\epsilon(i, t) = \alpha\sum_{k^{'}\in K\backslash k} |I_{tk^{'}}|d_{k_{\text{inf}}k_{\text{inf}}^{'}}^{-\tilde{\beta}},\ \alpha,\tilde{\beta}>0,$$
denoted as C-ILM M3 in Web Table~1. By allowing the between-cluster distances to vary over time, depending on which individuals are infectious, we might expect the spark function to better mimic between-cluster infections. However, the fact that the cluster centroid now varies over time does increase the computational cost. \par
Finally, to increase the flexibility of the model, we can consider making the effect of the number of infectious individuals on the between-cluster infectivity rate non-linear, using the spark function denoted as C-ILM M4 in Web Table~1: $$\epsilon(i, t) = \alpha\sum_{k^{'}\in K\backslash k} |I_{tk^{'}}|^{\delta}d_{kk^{'}}^{-\tilde{\beta}}\ \alpha,\tilde{\beta}>0, \delta\in\mathbb{R}.$$

\section{Dirichlet process mixture modelling for cluster identification}
\label{s:dpmm}
Parametric methods, such as Gaussian mixture models clustering via an expectation-maximization (EM) algorithm, rely on assumptions about the data's underlying distribution. These methods also typically require the number of clusters to be specified in advance, which can be a limitation when the true number of clusters is unknown. Bayesian nonparametric methods accommodate an infinite number of parameters, with complexity dynamically adapting to the data \citep{gershman2012tutorial}. One prominent such tool is the Dirichlet process (DP). A DP is a stochastic process yielding probability measures with total probability one, interpreted as distributions over a probability space $\Theta$ \citep{teh2010dirichlet}. Many intuitive representations have been proposed for constructing the DP, including the Chinese restaurant process~\citep{10.1007/BFb0099421, blackwell1973ferguson}, the stick-breaking process~\citep{sethuraman1994constructive}, and the P\'{o}lya Urn~\citep{blackwell1973ferguson}. Here, we consider clustering the population as a preparatory step before fitting the C-ILM using a Dirichlet Process Mixture Model (DPMM), which has been previously used in clustering epidemic data \citep[e.g.,][]{park2023spatio, wehrhahn2020bayesian}. Here, both spatial and temporal information are considered when clustering the population, via the DPMM method.
\subsection{Spatio-temporal distribution}
Given the cluster membership, $g_n$, of individual $n$, where $g_n = j, j\in \left\{1,\dots, M\right\}$ for an assumed finite number of $M$ clusters, the spatial location and time of infection can be modelled using the following spatial and temporal distributions, respectively. The spatial location of individual $n$, $\textbf{s}_n=(x_n, y_n)$, given $g_n = j$ is modelled using a bivariate Gaussian distribution:
\begin{equation*}
    \textbf{s}_n|g_n = j \sim \mathcal{N}\left(\binom{c_j^x}{c_j^y},\begin{pmatrix}
    \omega_x^2 & 0 \\
    0 & \omega_y^2
\end{pmatrix}\right),
\end{equation*}
where $\binom{c_j^x}{c_j^y}$ is the spatial mean vector of cluster $j$, and $\omega_x^2$, $\omega_y^2$ are the variance in the $x$ and $y$ directions, respectively, which are assumed to be homogeneous across clusters. The probability of the spatial location given cluster membership is: 
\begin{equation*}
    P(x_n, y_n|g_n = j) \propto \frac{1}{w_x w_y}\exp \left(-\frac{(x_n - c_j^x)^2}{2\omega_x^2} -\frac{(y_n - c_j^y)^2}{2\omega_y^2}\right).
\end{equation*}
Since infection times are discrete and we need to allow for those individuals that remain uninfected by the end of the epidemic, the commonly used Gaussian distribution is not suitable for modelling infection time. Thus, the infection time of individual $n$, $t_n$ given $g_n = j$ is modelled using a hurdle negative binomial distribution~\citep{cragg1971some}:
\begin{equation*}
    t_n|g_n = j \sim hNB(\theta_j, c_j^t, \phi),
\end{equation*}
where $\theta_j$ is the probability that cluster $j$ never gets infected (i.e. no individual moves to the infectious state throughout the observed epidemic), $c_j^t$ is the mean time of infection for all individuals in cluster $j$, and $\phi$ is the dispersion parameter, such that:
\begin{equation}\label{eq:conditional t}
     P(t_n|g_n = j) = \begin{cases} 
    \theta_j & \text{if } t_n = 0 \\
     (1-\theta_j)\frac{NB(t_n|c_j^{t},\phi)}{1-NB(0|c_j^{t},\phi)}& \text{if } t_n> 0 
     \end{cases}.
\end{equation}
We use the alternative parameterization of the negative binomial (NB) distribution \citep{Hilbe_2011}, given by
\begin{equation*}
    NB(t|\mu, \phi) = \binom{t+\phi-1}{t}(\frac{\mu}{\mu+\phi})^{t}(\frac{\phi}{\mu+\phi})^\phi,\ \mu,\phi\in \mathbb{R}^+,\ t\in \mathbb{N}
\end{equation*}
as in this form, $E(t) = \mu$ and $Var(t) = \mu + \frac{\mu^2}{\phi}$, where $\phi$ is the variance of the NB distribution, controlling the overdispersion scaled by $\mu^2$. We assume that the hurdle parameter $\theta_j$ and mean infection time $c_j^t$ are cluster-dependent, while $\phi$ is universal to all clusters. The full likelihood of $(\textbf{s}, t)$ can now be derived as follows:
\begin{equation*}
\begin{aligned}
    \mathcal{L}(\omega_x, \omega_y, \phi, \theta_{1:M},\textbf{c}_{1:M}^{x,y,t})&=\prod_{n=1}^{N}\mathcal{L}_n(\omega_x, \omega_y, \phi, \theta_{1:M},\textbf{c}_{1:M}^{x,y,t})\\
    &=\prod_{n=1}^N\sum_{j=1}^M P(g_n = j) P(x_n, y_n, t_n|g_n = j)\\
    & = \prod_{n=1}^{N}\sum_{j=1}^{M}\pi_j P(x_n, y_n|g_n = j)P(t_n|g_n=j)
\end{aligned}
\end{equation*}
where $\pi_j, j\in \left\{1,\dots, M\right\}$, is the latent class variable and $\textbf{c}^{x,y,t}$ is the cluster-dependent spatio-temporal mean $\textbf{c}_{1:M}^{x,y,t} = (c_{1:M}^x, c_{1:M}^y, c_{1:M}^t)$.
Note that ideally, we might want to use the ILM likelihood itself to inform cluster membership. However, doing so is computationally prohibitive for the reasons discussed in Section~\ref{s:discuss}.
\subsection{Dirichlet process mixture models}
\label{ss:dirichletprocess}
\textbf{Prior Distributions}\\
The cluster membership $g_{1:N}\in \left\{1,\dots, M\right\}$ for $N$ individuals can be modelled by a categorical distribution with latent class probabilities $\pi_{1:M}$
\begin{equation*}
    g_n\sim \text{Cat} (\pi_1,\dots, \pi_M).
\end{equation*}
The probabilities are defined by a stick-breaking process, which is widely used for constructing random weights~\citep{ferguson1973bayesian}:
\begin{equation}\label{stickbreak}
    \begin{aligned}
        &U_i\sim Beta(1,\gamma), \ \text{for } i = 1, \dots, M-1\\
        &U_M  = 1,\\
        &\pi_i = 
\begin{cases}
    U_1, & \text{if } i = 1, \\
    \prod_{j=1}^{i-1}(1-U_j)U_i, &  \text{if } i=2,\dots,M.
\end{cases}
    \end{aligned}
\end{equation}
where $\gamma$ is the rate parameter, and we define a hyperprior:
\begin{equation*}
    \gamma\sim \Gamma(1,2)
\end{equation*}
Although the stick breaking prior considers infinite $M$ in theory, we can approximate it with a reasonably large $M<\infty$ in practice. Here we use $M = 30$.\\\par
Before defining the cluster spatio-temporal means, we first standardize $\textbf{s}$ and $t$ to ensure they are on the same scale:
\begin{equation*}
    \begin{aligned}
        x_i &= t_{\text{min}}+(t_{\text{max}}-t_{\text{min}})\frac{x_i-x_{\text{min}}}{x_{\text{max}}-x_{\text{min}}}\\
    y_i &= t_{\text{min}}+(t_{\text{max}}-t_{\text{min}})\frac{y_i-y_{\text{min}}}{y_{\text{max}}-y_{\text{min}}}
    \end{aligned}
\end{equation*}
 
 The cluster spatio-temporal means $\textbf{c}_{1:M}^{x,y,t}$ are assumed to follow uniform distributions:
\begin{equation*}
    c_{m}^x, c_{m}^y, c_{m}^t \sim U(t_{\text{min}}, t_{\text{max}}), \quad m \in {1,\dots, M}
\end{equation*}
For the model variance parameters $\omega_x, \omega_y, \phi$:
\begin{equation*}
        \omega_x, \omega_y, \phi \sim \Gamma(1.5, 1) 
\end{equation*}
and for the hurdle parameters $\theta_{1:M}$
\begin{equation*}
    \theta_{1:M} \sim Beta(2, 2)
\end{equation*}
\textbf{Conditional Distributions and MCMC}\\
Before applying Gibbs sampling to sample from the posterior distribution, we first derive the conditional posterior distributions for: the variance-related parameters $\omega_x$, $\omega_y$, $\phi$; the cluster-dependent spatio-temporal means $\textbf{c}_{1:M}^{x,y,t}$ and hurdle parameters $\theta_{1:M}$; and the latent cluster membership variables $g_{1:N}$ and DPMM parameters $U_{1:M}$, $\pi_{1:M}$, $\gamma$. Detailed derivations can be found in the Web Appendix~A.\par
Conditional posterior distributions for variance-related parameters $\omega_x$, $\omega_y$, $\phi$ are as follows:
\begin{equation*}
\begin{aligned}
     P(\omega_x|\cdot)&\propto \mathcal{L}(\omega_x,\omega_y, \phi, \theta_{1:M},\textbf{c}_{1:M}^{x,y,t})P(\omega_x)\\
     P(\omega_y|\cdot)&\propto \mathcal{L}(\omega_x,\omega_y, \phi, \theta_{1:M},\textbf{c}_{1:M}^{x,y,t})P(\omega_y)\\
     P(\phi|\cdot)&\propto \mathcal{L}(\omega_x,\omega_y, \phi, \theta_{1:M},\textbf{c}_{1:M}^{x,y,t})P(\phi)
\end{aligned}
\end{equation*}
Conditional posterior distributions for cluster-dependent spatio-temporal means $\textbf{c}_{1:M}^{x,y,t}$ and hurdle parameters $\theta_{1:M}$ for $m = 1, \dots, M$ are as follows:
\begin{equation*}
\begin{aligned}
    P(c_m^x|\cdot)&\propto \prod_{n =1}^{N} \mathbb{I}_{\{g_n = m \}} P(x_n, y_n, t_n|g_n = m)P(c_m^x) \\
    & \propto \prod_{n =1}^{N} \mathbb{I}_{\{g_n = m \}} \exp{(-\frac{(x_n - c_m^x)^2}{2\omega_x^2})}
\end{aligned}
\end{equation*}
So $c_m^x\mid \cdot \sim \mathcal{N}(\bar{x}_m,\frac{\omega_x^2}{n_m})$, and by similar logic $c_m^y\mid \cdot \sim \mathcal{N}(\bar{y}_m, \frac{\omega_y^2}{n_m})$ where $\bar{x}_m$ and $\bar{y}_m$ refers to the mean of $x_n$ and $y_n$ that belongs to cluster $m$, and $n_m$ refers to the number of individuals in cluster $m$. Further, 
\begin{equation*}
    \begin{aligned}
    P(c_m^t|\cdot)&\propto \prod_{n =1}^{N} \mathbb{I}_{\{g_n = m \}} P(x_n, y_n, t_n|g_n = m)P(c_m^t) \\
    & \propto \prod_{n =1}^{N} \mathbb{I}_{\{g_n = m \}} P(t_n|g_n = m)
\end{aligned}
\end{equation*}
\begin{equation*}
\begin{aligned}
     P(\theta_m|\cdot) &\propto \prod_{n =1}^{N} \mathbb{I}_{\{g_n = m \}} P(x_n, y_n, t_n|g_n = m)P(\theta_m) \\
     &\propto \prod_{n =1}^{N} \mathbb{I}_{\{g_n = m \}} P(t_n|g_n = m)P(\theta_m)
\end{aligned}
\end{equation*}
where $P(t_n | g_n = m)$ refers to Equation (\ref{eq:conditional t}). \\
The conditional posterior distribution for the latent cluster membership variables $g_{1:N}$ are given by:
\begin{equation*}
\begin{aligned}
    P(g_n = m|\cdot) &\propto \pi_m P(x_n, y_n, t_n|g_n = m)\\
    &\propto \pi_m P(x_n, y_n |g_n = m)P(t_n|g_n = m)
\end{aligned}
\end{equation*}
for $n = 1,\dots, N$ and $m = 1, \dots, M$. \\
The conditional posterior distributions for the DPMM parameters $U_{1:M}$, $\pi_{1:M}$ and $\gamma$ are given by:
\begin{equation*}
    \begin{aligned}
        P(U_m|\cdot)&\propto P(g_{1:N}|U_m)P(U_m|\gamma)
    \end{aligned}
\end{equation*}
So $$U_m \mid \cdot\sim Beta (n_m + 1, \gamma+\sum_{j = m+1}^M n_j)$$
for $m = 1,\cdots, M$, where $n_m$ and $n_j$ are the number of individuals assigned to to cluster $m$ and $j$.
Finally,
\begin{equation*}
    \begin{aligned}
    P(\gamma|U_{1:M})&\propto P(U_{1:M}|\gamma)P(\gamma)\\
\end{aligned}
\end{equation*}
So $$\gamma\mid U_{1:M}\sim \Gamma(M, 2-\sum_{m=1}^{M-1}\log(1-U_m) )$$
As $\pi_{1:M}$ are deterministically determined by $U_{1:M}$, the update for $\pi_{1:M}$ can be calculated using Equation (\ref{stickbreak}).
Then we can formulate the Gibbs sampling process into the DPMM-C-ILM MCMC algorithm, which is Algorithm 1 in Web Appendix~B. \par
The posterior samples include each individual’s cluster membership and the corresponding cluster centroids.
However, only posterior point estimates are passed to the second stage of the study, which involves analyzing the epidemic data using the CILM described in Section~\ref{s:model}. Cluster posterior uncertainty could possibly be propagated through to the CILM analysis stage, so as to jointly model the clustering and epidemic dynamics. However, this would come with a large computational cost. This possible propagation of uncertainty from the first stage to the second is discussed further in Web Appendix C.

\section{Simulation study}\label{s:simulation}
\subsection{Simulation methodology}\label{ss:simulation_methodology}
A simulation study was conducted to explore how well the C-ILMs performed under different scenarios. Spatial locations of individuals in the population were generated under three spatial settings, described in more detail below: completely spatially random (CSR), low-variance clustered, and high-variance clustered. In each clustered scenario, $K = 3, 5, \text{and }8$  clusters were simulated. Therefore, in total we have seven spatial scenarios, and $m = 10$ populations containing $n = 100$ individuals each were generated for each spatial scenario. 
Epidemics were simulated from the basic SILM using Equation (\ref{eq:basicSILM}) with $\epsilon(i, t) = 0$. The epidemics were fitted under the basic SILM and the corresponding spark functions described in Section \ref{ss:spark} and Web Table~1. Both the DPMM clustering method of Section \ref{s:dpmm} and $K$-means clustering were considered as means of defining the clusters. $K$-means clustering is a centroid-based clustering algorithm that iteratively assigns individuals to the nearest centroids, and updates the centroids based on the spatial mean of the assigned individuals \citep{macqueen1967some}.\par
\noindent\textbf{Data Generation}. In the CSR populations, spatial coordinates for each individual in the population were randomly sampled from independent uniform distributions, $x, y\sim \mathcal{U}(0, 30)$. In clustered populations, spatial coordinates were randomly sampled from a bivariate Gaussian distribution: $$x_k, y_k \sim \mathcal{N}(\mu_k, \Sigma)$$ where $\mu_{k}=\begin{bmatrix}
\mu_{k1} \\
\mu_{k2}
\end{bmatrix}$ is the mean of cluster $k$, and $\mu_{k1}$, $\mu_{k2}\sim \mathcal{U}(0,30)$. In high-variance clustered populations $\Sigma = \begin{bmatrix} 8 & 0 \\ 0 & 8 \end{bmatrix}$, while in low-variance clustered populations $\Sigma = \begin{bmatrix} 3 & 0 \\ 0 & 3 \end{bmatrix}$.\par
Epidemics were generated within an SIR compartmental framework for a total of $t_{\text{max}}=31$ discrete time points. The epidemic starts at time $t=0$, with a single individual selected at random to initialize epidemics, and the infectious period is 3 days for each individual. Throughout, the baseline susceptibility $\alpha$ was set to $\alpha = 0.8$, and the spatial parameter was set to $\beta = 2$. Web Figure~1 shows the 10 resulting epidemic curves generated under each spatial scenario.\par
\noindent \textbf{Prior Distribution and Model Assessment}. For all models, weakly informative independent priors for epidemic parameters $\alpha$, $\beta$ and $\tilde{\beta}$ were used, such that: $\alpha \sim \Gamma(1.5 ,1)$ and $\beta, \tilde{\beta}\sim \Gamma(2 , 3)$.
The basic SILM and each C-ILMs were fitted to each data set. The Watanabe-Akaike information criterion (WAIC) was used for model comparison. To further evaluate model fit, we considered the posterior predictive distribution (PPD) of the epidemic curve. Specifically, here we used the incidence curve: 
\begin{equation*}\label{epicurve}
    \mathscr{C} = \{|I(t+1)\backslash I(t)|\}^{t_\text{max}}_{t=1}.
\end{equation*}
The posterior predictive distribution of daily incidence, $\mathscr{C}$, was estimated by simulating 100 parameter sets from the posterior distributions estimated via MCMC. We considered two different PPD forecasts: complete-case PPD and PPD forecasting. Complete-case PPD utilized all the epidemic data to estimate posteriors and then simulated the entire epidemic, while PPD forecasting regenerated the epidemic curves from time $t=5$, based on models fitted to the first 5 time points only. 95\% highest posterior density intervals (HPDIs) were also calculated for $\mathscr{C}$ at each time point.

\subsection{Results}
\begin{figure}[H]
  \centering \includegraphics[width=0.7\textwidth]{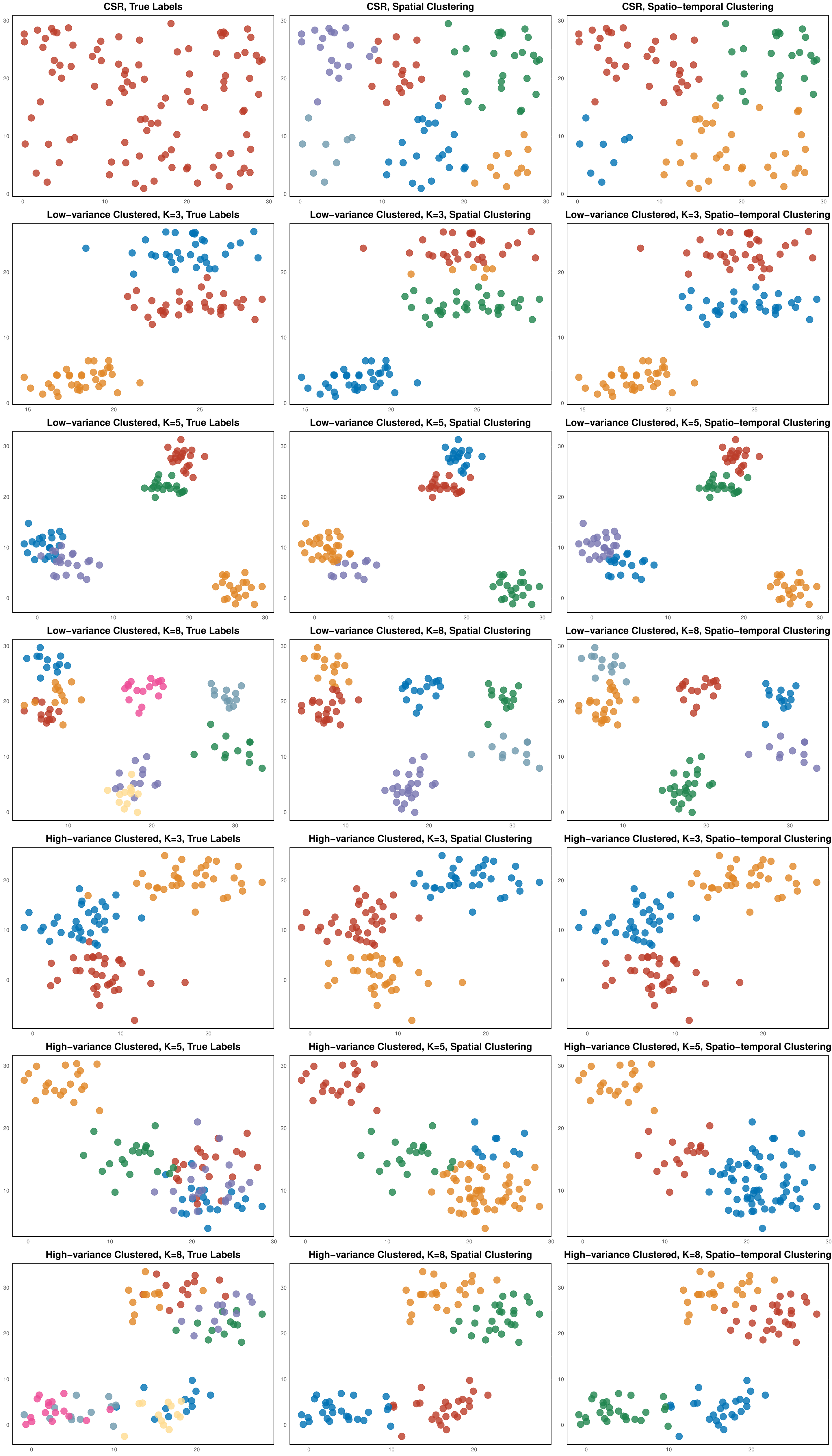}
  \caption{Clustering results for spatial and spatio-temporal data in seven spatial scenarios. Different colours represent different clusters. The notation `True Label' refers to the actual cluster labels assigned to each individual. `Spatial Clustering' indicates the clustering based on spatial data and `Spatio-temporal Clustering' denotes the result that takes both spatial and temporal data into account.}
  \label{fig:clustering_1}
\end{figure}

\begin{figure}[htbp]
  \centering
  \begin{minipage}{0.396\textwidth} 
    \centering
    \includegraphics[height=16.5cm,
      width = \linewidth, 
      keepaspectratio,
      valign=c]{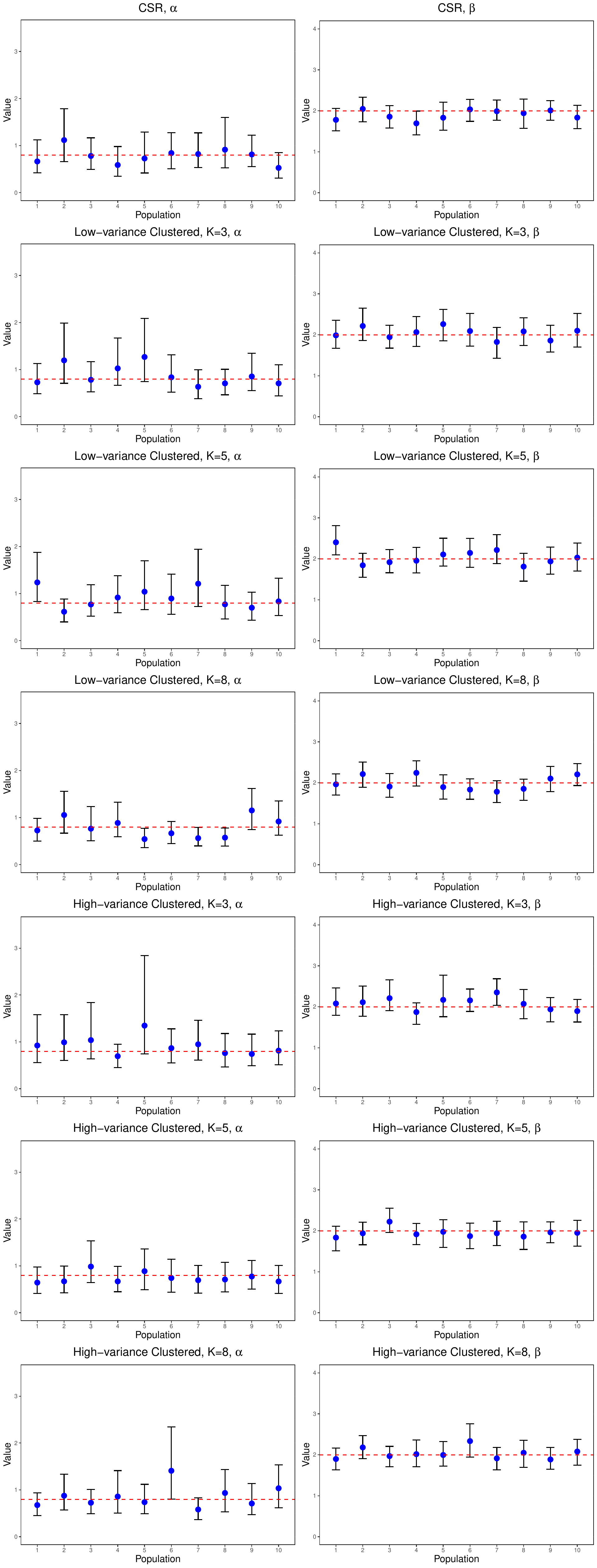} 
    \captionof{subfigure}{basic SILM} 
    \label{fig:PE_basic}
  \end{minipage}
  \hfill
  \begin{minipage}{0.594\textwidth}
    \centering
    \includegraphics[height=16.5cm,               
      width= \linewidth,
      keepaspectratio,
      valign=c ]{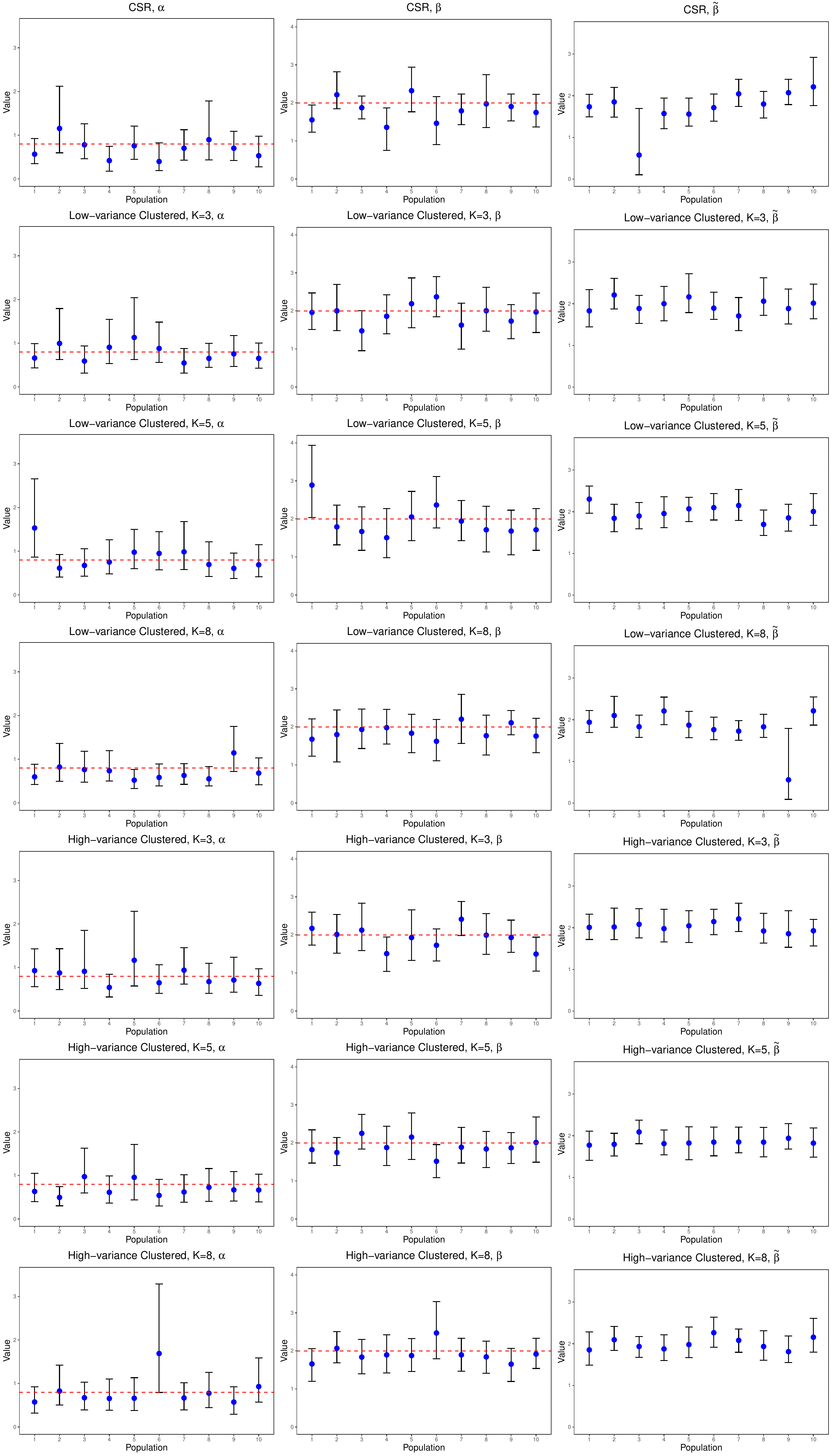}
    \captionof{subfigure}{M2}
    \label{fig:PE_M2}
  \end{minipage}
  \caption{95\% HPDIs and posterior medians (blue dots) for basic SILM and M2 parameters. Red horizontal lines represent the true parameter values.} 
  \label{fig:PE_basic&M2}
\end{figure}

\noindent\textbf{Clustering}. As described, we applied $K$-means algorithm with $K=3, 5, 8, \text{or }10$ to generate the corresponding number of clusters. The DPMM clustering algorithm was applied to spatial data only $(x_n, y_n)$, as well as spatio-temporal data ($x_n, y_n, t_n$), using MCMC chains of length 2000. Figure~\ref{fig:clustering_1} shows the identified clusters, colour-coded according to the cluster to which each individual belongs, for one typical epidemic under each spatial scenario. The $K$-means results are omitted here for simplicity. It is observed that when considering the infection time ($t_n$), uneven clusters are more likely to form, with some clusters containing very few individuals (e.g., in CSR and high-variance clustered $K=5$). Additionally, spatio-temporal DPMM clustering tends to generate fewer clusters compared to spatial-only clustering (in CSR, low-variance clustered $K=3$ and high-variance clustered $K=5$ scenarios). This seemingly occurs because spatio-temporal clustering groups individuals who are infected around the same time into the same cluster, which is reasonable as individuals are more likely infected by the individuals who recently become infected/infectious.
\noindent\textbf{C-ILM Performance}. The MCMC chains for the C-ILMs with no spark function, constant spark function, and M1 did not mix well. Therefore,  we exclude these models from further discussion here. Figure~\ref{fig:PE_basic&M2} displays the 95\% HPDIs and posterior medians for the parameter estimates obtained under basic SILM and model M2. The posterior estimates under model M3 and M4 can be found in the Web Figures~2 and~3, respectively. Across all populations, the parameter estimation for $\alpha$ and $\beta$ was successful, with the 95\% HPDIs capturing the true parameter values in most cases. However, in models containing the $\tilde{\beta}$ between-cluster spatial parameter, the 95\% HPDIs of other parameters are often wider than their equivalents under the SILM.\par

\begin{figure}[htbp]
  \centering
\includegraphics[width=0.7\textwidth]{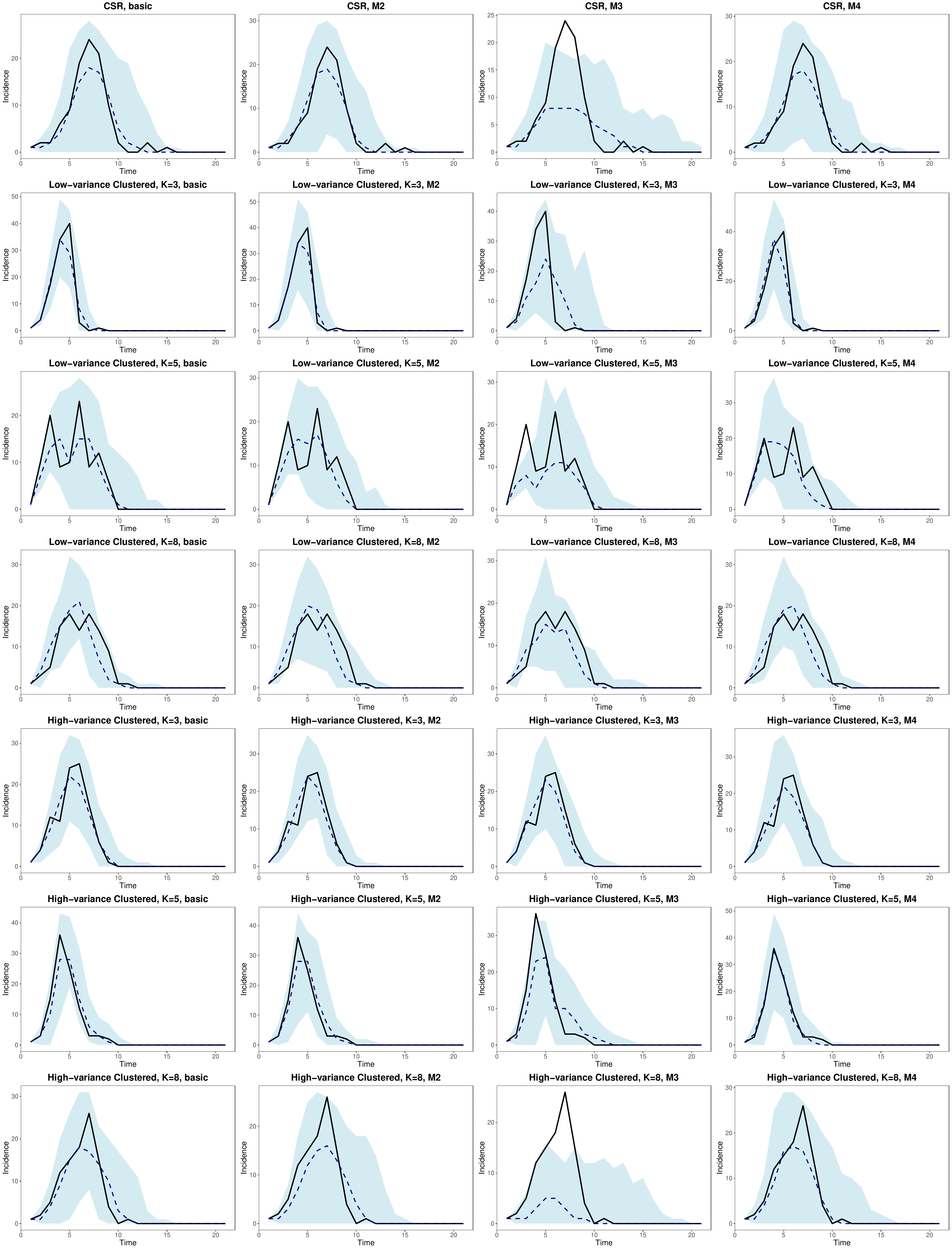}
  \caption{95\% HPDIs (blue shadows) for the epidemic curve of one representative epidemic in each scenario, where the posterior medians are shown as dashed blue lines, the posterior medians are shown as blue solid line. The true epidemic curves are shown with solid black lines.}
  \label{fig:complete-case PPD}
\end{figure}

We assessed the performance of the basic SILM and C-ILM models across various population settings by analyzing the posterior predictive distributions (PPDs) of the epidemic curves. Figure~\ref{fig:complete-case PPD} illustrates the results for one typical epidemic from each of the seven spatial scenarios. In general, the 95\% HPDIs of the epidemic curves under the basic SILM, M2 and M4 capture the true curves well, indicating a good fit. M3 tends to exhibit the poorest performance, particularly in scenarios with less clear patterns of spatial clustering.\par
We also assessed the forecasting capability of the basic SILM and C-ILMs by utilizing the posterior estimates to generate epidemics across various time points from day 5 onwards. Web Figure~4 illustrates the forecasts for one typical epidemic under each scenario. Once again, in general, the 95\% HPDIs of the forecasts for the basic SILM and C-ILMs adequately captured every epidemic curve in all spatial scenarios, except under M3 with $K=5$ or 8 in low-variance clustered population.\par
We also evaluated the performance of different C-ILMs under the DPMM and $K$-means clustering algorithm using the WAIC. The results are summarized in Table~\ref{tab:WAICM234}. The lowest WAIC values were achieved consistently by using M2 spark function across all spatial scenarios. Thus, based on the WAIC and the PPD of the epidemic curve, we conclude that M2 provides a superior fit compared to the other C-ILMs. Comparing the clustering algorithms under model M2, the DPMM consistently achieves the lowest WAIC values across all spatial scenarios, with the exception of the high-variance clustered population at $K=8$. Under the second best C-ILM, M4, the DPMM also outperforms all variations of the $K$-means clustering algorithm across every spatial scenario, further supporting the effectiveness of the DPMM as a clustering method.
\par
\begin{table}[H]
    \centering
    \caption{WAIC values for different clustering methods using M2, M3, and M4 across all scenarios, with the minimum value highlighted in bold.}
    \renewcommand{\arraystretch}{1.5} 

    \begin{minipage}{\textwidth}
        \centering
        \text{(a) WAIC values for different clustering methods using M2 in all scenarios}
        \begin{adjustbox}{width=\textwidth}
        \begin{tabular}{cccccc}
            \hline
              \textbf{M2}& $K\text{-means}, K=3$ & $K\text{-means}, K=5$ & $K\text{-means}, K=8$ & $K\text{-means}, K=10$ & DPMM \\
            \hline
            CSR & 3667.43 & 3659.94 & 3661.13 & 3651.36 & \textbf{3626.89} \\
            Low-variance Clustered, $K=3$ & 1776.42 & 1809.42 & 1840.92 & 1864.17 & \textbf{1764.32} \\
            Low-variance Clustered, $K=5$ & 2148.4 & 2147.97 & 2186.32& 2179.41 & \textbf{2145.88} \\
            Low-variance Clustered, $K=8$ & 2591.37 & 2619.84 & 2599.41 & 2644.42 & \textbf{2585.66} \\
            High-variance Clustered, $K=3$ & 2549.06 & 2582.97 &  2624.35 & 2606.2 & \textbf{2540.47} \\
            High-variance Clustered, $K=5$ & 2852.82 & 2875.48 & 2889.37 & 2883.58 & \textbf{2832.42} \\
            High-variance Clustered, $K=8$ & \textbf{3111.48} & 3126.9 & 3160.2 & 3154.17 & 3115.14 \\
            \hline
        \end{tabular}
        \end{adjustbox}
    \end{minipage}
    
    \vspace{1cm} 

    \begin{minipage}{\textwidth}
        \centering
        \text{(b) WAIC values for different clustering methods using M3 in all scenarios}
        \begin{adjustbox}{width=\textwidth}
        \begin{tabular}{cccccc}
            \hline
              \textbf{M3}& $K\text{-means}, K=3$ & $K\text{-means}, K=5$ & $K\text{-means}, K=8$ & $K\text{-means}, K=10$ & DPMM \\
            \hline
            CSR & \textbf{4278.11} & 5070.71 & 5978.28 & 7337.16 & 4508.57 \\
            Low-variance Clustered, $K=3$ & 2165.24 & 2274.94 & 2923.98 & 3018.15 & \textbf{2096.41} \\
            Low-variance Clustered, $K=5$ & \textbf{2383.64} & 3047.92 & 4129.04 & 4887.66 & 2409.23 \\
            Low-variance Clustered, $K=8$ & \textbf{2979.3} & 3394.54 & 4352.81 & 5015.55 & 3678.08 \\
            High-variance Clustered, $K=3$ & 2720.81 & 3032.08 &  4130.52 & 4834.84 & \textbf{2664.48} \\
            High-variance Clustered, $K=5$ & \textbf{3117.05} & 3826.54 & 4910.92 & 5849.53 & 3117.14 \\
            High-variance Clustered, $K=8$ & \textbf{3490.11} & 4287.82 & 5957.07 & 6575.82 & 3736.92 \\
            \hline
        \end{tabular}
        \end{adjustbox}
    \end{minipage}
    
    \vspace{1cm} 

    \begin{minipage}{\textwidth}
        \centering
        \text{(c) WAIC values for different clustering methods using M4 in all scenarios}
        \begin{adjustbox}{width=\textwidth}
        \begin{tabular}{cccccc}
            \hline
              \textbf{M4}& $K\text{-means}, K=3$ & $K\text{-means}, K=5$ & $K\text{-means}, K=8$ & $K\text{-means}, K=10$ & DPMM \\
            \hline
            CSR & 3689.30 & 3672.25 & 3666.45 & 3659.95 & \textbf{3637.53} \\
            Low-variance Clustered, $K=3$ & 1794.44 & 1818.77 & 1846.45 & 1876.17 & \textbf{1777.34} \\
            Low-variance Clustered, $K=5$ & 2158.27 & 2160.72 & 2198.95 & 2197.05 & \textbf{2151.18} \\
            Low-variance Clustered, $K=8$ & 2606.62 & 2630.86 & 2603.20 & 2657.83 & \textbf{2594.79} \\
            High-variance Clustered, $K=3$ & 2570.59 & 2582.70 &  2646.60 & 2610.83 & \textbf{2547.69} \\
            High-variance Clustered, $K=5$ & 2861.79 & 2872.88 & 2896.04 & 2887.06 & \textbf{2843.82} \\
            High-variance Clustered, $K=8$ & 3130.05 & 3135.51 & 3171.62 & 3166.38 & \textbf{3128.81} \\
            \hline
        \end{tabular}
        \end{adjustbox}
    \end{minipage}

    \label{tab:WAICM234}
\end{table}

\section{Application: 2001 U.K. foot and mouth disease}
\label{s:application}
\subsection{Data analysis and methodology}
In this section, we applied the C-ILMs to data from the 2001 U.K. foot and mouth disease (FMD) epidemic, an outbreak that predominantly affected sheep and cattle farms. We focussed on a subset of observations from the county of Cumbria in the north-west of England (Web Figure~5(a)), a region with a high concentration of farms and a relatively high incidence of disease, following \citet{DeethDeardon+2013+75+93}. The data contains geographical information for 1,177 livestock farms (cattle and sheep), as well as infection and removal times for each farm. The first infection in the region in our dataset was recorded on day 21 of the epidemic. Various control strategies were used to bring the disease under control, most predominantly a mass cull of animals on infected and ``at-risk'' farms \citep{kitching2005review}. We fit the model to the data from day 30 to 50 (8-28 March 2001) of the epidemic. A simple spatial SEIR compartmental model was fitted to the data. Under this model, the transition from susceptible to exposed was modelled using the basic SILM:
\begin{equation}
    P(i,t) = 1-\exp{\left[-\alpha\sum_{j\in I(t)}d_{ij}^{-\beta}\right]},\ \alpha,\beta>0
\end{equation}
The transition time from exposed to infectious was assumed to be known and fixed at $\gamma_E = 5$ days \citep{DeethDeardon+2013+75+93}. The C-ILM with the M2 spark function, the best performing C-ILM in the simulation study from the previous section, was also fitted to the data:
\begin{equation}
    P(i,t) = 1-\exp{\left[-\alpha\left(\sum_{j\in I_k(t)}d_{ij}^{-\beta}+\sum_{k^{'}\in K\backslash k}|I_{tk^{'}}|d_{kk^{'}}^{-\tilde{\beta}}\right)\right]},\ \alpha,\beta,\tilde{\beta}>0
\end{equation}
The infection period was also assumed known and fixed, here at $\gamma_I = 4$ days, unless animals on the farm were culled earlier as part of the control policy, in which case the farm would transition to the removed state on the culling date.\par 
We used the same prior and MCMC settings as in the simulation study, running for 2,000 iterations. Convergence was assessed using trace plots, the Gelman-Rubin statistic, and the stability of cluster assignments. The cluster assignments, based on the DPMM clustering algorithm, are shown in Web Figure~5(b), where different colours represent different clusters. The algorithm grouped the farms into 27 distinct clusters. We also use the $K$-means clustering algorithm purely on spatial location data, with $K = 10, 20, 27, \text{and }30$. 
\subsection{Results}
The posterior estimates along with their 95\% HPDIs are presented in Table~\ref{tab:PE_FMD}, under the basic SILM, C-ILM with $K$-means clustering ($K = 10, 20, 27, 30$), and the C-ILM with the DPMM clustering method. We observed that the $\alpha$ estimates were consistent across all models, averaging around 0.007. However, the $\beta$ estimates varied somewhat. As the number of clusters increased, the $\beta$ estimates tended to decrease, while the $\tilde{\beta}$ estimates remained relatively consistent across all composite models. 
\begin{sidewaystable}
\setlength{\belowcaptionskip}{0.2cm}
\caption{Parameter posterior median estimates and 95\% HPDIs for basic SILM and C-ILMs fit to the FMD Cumbria dataset.}
\centering

\scriptsize
\begin{tabularx}{\textwidth}{c>{\centering\arraybackslash}X>{\centering\arraybackslash}X>{\centering\arraybackslash}X>{\centering\arraybackslash}X>{\centering\arraybackslash}X>{\centering\arraybackslash}X}
\toprule
\multirow{2}{*}{Parameter} & \multicolumn{2}{c}{basic SILM} & \multicolumn{2}{c}{C-ILM, $K\text{-means}, K=10$} & \multicolumn{2}{c}{C-ILM, $K\text{-means}, K=20$} \\ 
\cmidrule(lr){2-3} \cmidrule(lr){4-5} \cmidrule(lr){6-7}
                           & Estimate & 95\% HPDI & Estimate & 95\% HPDI & Estimate & 95\% HPDI \\ 
\midrule
$\alpha$                          & 0.006        & (0.004, 0.008)        & 0.006        & (0.004, 0.008)        & 0.007        & (0.005, 0.010)        \\
$\beta$                       & 1.200        & (1.041, 1.344)        & 0.745        & (0.432, 1.039)        & 0.772        & (0.481, 1.064)        \\ 
$\tilde{\beta}$                    &         &         & 1.298        & (1.068, 1.475)        & 1.351        & (1.152, 1.554)        \\
\bottomrule
\end{tabularx}

\vspace{0.5cm} 

\begin{tabularx}{\textwidth}{c>{\centering\arraybackslash}X>{\centering\arraybackslash}X>{\centering\arraybackslash}X>{\centering\arraybackslash}X>{\centering\arraybackslash}X>{\centering\arraybackslash}X}
\toprule
\multirow{2}{*}{Parameter} & \multicolumn{2}{c}{C-ILM, $K\text{-means}, K=27$} & \multicolumn{2}{c}{C-ILM, $K\text{-means}, K=30$} & \multicolumn{2}{c}{C-ILM, DPMM} \\ 
\cmidrule(lr){2-3} \cmidrule(lr){4-5} \cmidrule(lr){6-7}
                           & Estimate & 95\% HPDI & Estimate & 95\% HPDI & Estimate & 95\% HPDI \\ 
\midrule
$\alpha$                          & 0.007        & (0.005, 0.010) & 0.007        & (0.004, 0.009)        & 0.007        & (0.004, 0.010)        \\
$\beta$                       & 0.758        & (0.435, 1.112)        & 0.541        & (0.128, 0.891)        & 0.361        & (0.047, 0.696)        \\ 
$\tilde{\beta}$                    & 1.319        & (1.109, 1.532)        & 1.319        & (1.112, 1.527)        & 1.467        & (1.230, 1.665)        \\
\bottomrule
\label{tab:PE_FMD}
\end{tabularx}
\end{sidewaystable}
Due to the well-known difficulty in simulating the culling-based control strategy used in the 2001 outbreak, it is difficult to compare these models using posterior predictive epidemic curves. However, based on the WAIC values in Web Table~2, the C-ILM with DPMM clustering algorithm has the smallest WAIC value of 1963.946, which is lower than the basic SILM's WAIC value of 1993.028. This suggests that the C-ILM with DPMM clustering provides a better fit to the 2001 UK FMD data compared to the basic SILM. Moreover, the computation time for M2 is significantly reduced compared to the basic SILM. For example, to run 2000 iterations on a macOS 11(Big Sur) Version 11.6, with a 1.1 GHz Dual-Core Intel Core i3 processor, the basic SILM takes 2649.583 seconds, whereas the C-ILM only requires 517.709 seconds.
\section{Discussion}
\label{s:discuss} 
Computational challenges in inference for ILMs are a well-known challenge. By applying the composite method, the population is divided into spatial subsets with low interaction, significantly reducing the degree of individual-individual transmission that needs to be accounted for in the model. The probability of infection between clusters can then be approximated using a well-chosen spark function. Within the C-ILM framework, we propose a DPMM clustering algorithm to partition the population, assuming static spatial locations that follow a bivariate Gaussian distribution and infection times that follow hurdle negative binomial distributions. We have demonstrated the effectiveness of these methods through both simulated and real-world data, showing that the C-ILMs and the DPMM clustering algorithm are useful for both parameter estimation and epidemic forecasting.\par
Ideally, we might want to use the C-ILM likelihood to inform the clustering processes. However, attempts to do this proved computationally untractable. Thus, the DPMM is applied using separate modelling components for the observed spatial locations and infection times, with a view to approximating the full C-ILM likelihood. Consequently, the DPMM clustering serves primarily as a computationally efficient means of partitioning the population, rather than a model-based representation of transmission heterogeneity. \par
Again for computational reasons, posterior uncertainty from the DPMM clustering is ignored in the C-ILM analysis in which we condition on the point estimates of the cluster memberships. However, we can consider how the posterior uncertainty from the first-stage DPMM clustering might propagate to the second-stage C-ILM analysis. This is shown in the Web Appendix C. Our analysis indicates that using the full set of posterior samples compared to using summary statistics, specifically the posterior mode of cluster memberships and the posterior median of cluster centroids, results in minimal differences in parameter estimates and posterior predictive distributions of incidence.\par
We also recorded the computation time that might be saved using a C-ILM to analyze data in a moderately sized population ($N = 1000$). Under a typical simulation study scenario, models M2 and M3 required about one-third of the time (462.74s and 470.96s, respectively) compared to the basic SILM (1206.11s), while M4 took roughly three-fifths of that time (736.62s). These results suggest that C-ILMs significantly improve computational efficiency, and this improvement will tend to be more dramatic as $N$ increases. Based on both WAIC values and computational speed, our results suggest that M2 provides a reasonable default option for its superior parameter estimation, forecasting ability, and reduced computation time.\par
Although the composite method was introduced primarily as computationally efficient approximation of the existing ILM framework, there may be many situations where the composite version of the ILM may lead to an improvement in model performance. We see this in the FMD example where the model with the smallest WAIC is the DPMM clustered C-ILM. This could occur because clustering might capture some real physical phenomena. For example, spatial and/or spatio-temporal clustering may occur in reality due to some physical barrier (e.g. river, land, otherwise unsuitable for the population). Under the basic SILM, transmission across such barriers is assumed unhindered (except for the distance itself), whereas the C-ILM can allow for this by impeding or reducing cross-barrier transmission. This also raises the possibility of using the C-ILM to detect heterogeneity in the population; for example, by allowing model parameters to vary between clusters. The potential for improvement of modelling through the composite approach is something that should be explored further. \par

Future research could also explore applying C-ILM to the geographically dependent ILM (GD-ILM) proposed by \citet{mahsin2022geographically}, which incorporates individual-level spatial location, spatially varying regional-level risk factors (e.g., socioeconomics, environment), and unobserved spatial structure into the susceptibility function. Currently, regions have been assumed known in these models, but could be defined by the clustering algorithm. The susceptibility function could vary by cluster, accounting for geographically dependent information and enhancing the model's ability to capture spatial heterogeneity in infection rates, leading to a more accurate fit to epidemic data. Another potential application is integrating C-ILM into the behavioural change ILM (BC-ILM), a framework that incorporates individual-level information and behavioural change effects modeled by ``alarm" functions \citep{ward2023bayesian, ward2025framework}.  Incorporating behavioural change effects would enable ILMs to better reflect real-world epidemics, accounting for changes in transmission-mitigating behaviour change in response to the outbreak. Then once again, the alarm function could be allowed to vary according to the cluster, detecting spatial heterogeneity in risk-avoiding behavioural change.

\section*{Acknowledgements}
This work was supported by funding from the Natural Sciences and Engineering Research Council of Canada (NSERC) Discovery Grants of Deardon and Deeth, University of Calgary Eyes High Doctoral Scholarship and Faculty of Graduate Studies Doctoral Scholarship. None of these funding agencies had direct involvement with the study.
\section*{Supporting Information}

Web Appendix, referenced in Section~\ref{ss:spark},~\ref{ss:dirichletprocess},~\ref{s:simulation} and~\ref{s:application} is provided alongside this paper. The code and data in this article are available at \url{https://github.com/YiraoZ/CILM}.












\end{document}